\documentclass[%
 reprint,
 amsmath,amssymb,
 aps,
 prl,
]{revtex4-1}

\usepackage{graphicx}
\usepackage{dcolumn}
\usepackage{bm}
\usepackage{titlesec}
\usepackage{color}

\begin{document}


\title{Exploring Event Horizons and Hawking Radiation through Deformed Graphene Membranes}

\author{Tommaso Morresi,$^{1,2}$ 
Daniele Binosi,$^{1}$ 
Stefano Simonucci,$^{3}$
Riccardo Piergallini,$^{3}$ 
Stephan Roche,$^{4,5\ast}$
Nicola M. Pugno,$^{2,6,7}$ 
Simone Taioli$^{1,8\ast}$\\~\\}
\affiliation{%
  $^{1}$European Centre for Theoretical Studies in Nuclear Physics and Related Areas (ECT*-FBK), Trento, Italy\\
$^{2}$Laboratory of Bio-Inspired \& Graphene Nanomechanics -- Department of Civil, Environmental and Mechanical Engineering, University of Trento, Italy\\
$^{3}$School of Science and Technology, University of Camerino, Camerino, Italy\\
$^{4}$ Catalan Institute of Nanoscience and Nanotechnology (ICN2),
CSIC and BIST, Campus UAB, Bellaterra, Barcelona, Spain\\
$^{5}$ICREA, Institucio Catalana de Recerca i Estudis Avancats, Barcelona, Spain  \\
$^{6}$Ket-Lab, Edoardo Amalfi Foundation, Rome, Italy\\
$^{7}$School of Engineering and Materials Science, Queen Mary University of London, UK \\
$^{8}$Trento Institute for Fundamental Physics and Applications (TIFPA-INFN), Trento, Italy 
}%

\begin{abstract}
Analogue gravitational systems are becoming an increasing popular way of studying the behaviour of quantum systems in curved spacetime. Setups based on ultracold quantum gases in particular, have been recently harnessed to explore the thermal nature of Hawking's and Unruh's radiation that was theoretically predicted almost 50 years ago. For solid state implementations, a promising system is graphene, in which a link between the Dirac-like low-energy electronic excitations and relativistic quantum field theories has been unveiled soon after its discovery. Here we show that this link extends to the case of curved quantum field theory when the graphene sheet is shaped in a surface of constant negative curvature, known as Beltrami's pseudosphere. Thanks to large-scale simulations, we provide numerical evidence that energetically stable negative curvature graphene surfaces can be realized; the ratio between the carbon-carbon bond length and the pseudosphere radius is small enough to allow the formation of an horizon; and the associated Local Density Of States evaluated at horizon's proximity has a thermal nature with a characteristic temperature of few tens of Kelvin. Such findings pave the way to the realization of a solid-state system in which the curved spacetime dynamics of quantum many body systems can be investigated.
\end{abstract}
\maketitle               

Quantum mechanics and general relativity are the most successful theories of modern physics. Most of the predicted exotic phenomena, from the weirdness of quantum entanglement to the existence of black holes have been experimentally tested and verified. On the other hand, a serious difficulty remains to merge those two fundamental theories in a single framework, which, in turn, makes it extremely challenging to obtain firm theoretical predictions. \\
\indent One remarkable exception is the discovery by Hawking that, from a quantum mechanical point of view, black holes are not completely black~\cite{Hawking:1974rv}: they emit `Hawking radiation' consisting of photons, neutrinos and, to a lesser extent, all sorts of massive particles. However, direct detection of this radiation, which is thermal in nature, seems beyond the experimental reach: Hawking radiation is in fact predicted to be proportional to the inverse of the black hole mass, which, for the smallest observed black hole, implies $T=60$~nK, i.e., 9 orders of magnitude smaller than the current cosmic microwave background  temperature. \\
\indent On the other hand, so-called black hole analogues, first proposed by Unruh~\cite{Unruh:1980cg}, are rapidly turning from promising to consolidated avenues in the study of various thermodynamics aspects. This is particularly true for sonic analogues built from ultracold gases~\cite{Jacobson:1998ms,Garay:1999sk,Barcelo:2001ca,Giovanazzi:2004xx,Balbinot:2007de,Carusotto:2008ep,Macher:2009nz,Recati:2009xx,Larre:2011mq,Steinhauer:2015ava,Jacobson:1998ms}, for which not only Unruh-\cite{Hu:2018psq} and Hawking-like~\cite{Steinhauer:2015saa} radiation has been experimentally observed, but, in the latter case, its correlation spectrum shown to be thermal and with a temperature given by the system's surface gravity~\cite{deNova:2018rld}, thus vindicating Hawking's predictions. \\
\indent The state-of-the-art of solid-state black hole analogues is, on the other hand, at a less advanced stage~\cite{Franz:2018cqi}. Indeed, while all current experimental approaches face major challenges mainly related to material synthesis and device fabrication, in the last couple of years key conceptual advances have been achieved; thus, there are now hopes for some of the fundamental questions to be addressed in condensed matter systems too, especially in connection to the implementation of the Sachdev--Ye--Kitaev model~\cite{Sachdev:1992fk,Kitaev:2015} and its potential to holographically realize quantum black holes. \\
\begin{figure*}[!t]
	\includegraphics[scale=0.5]{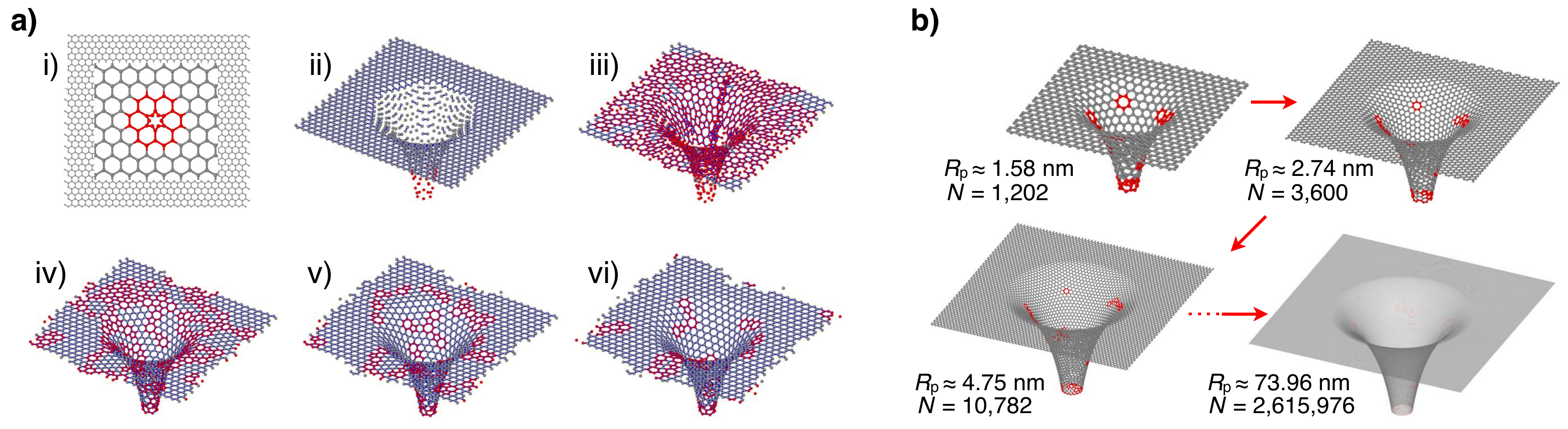}
	\caption{\label{fig:1}Tiling the hyperbolic plane by three-coordinated tessellations realized by an all-$sp^2$ carbon-based net. a) Optimization of a carbon pseudosphere containing $N=1626$ atoms with radius $R_\mathrm{p}=18.26$ \AA. Panels from i to vi represent different optimization stages, consisting of trial bond switch--twist moves. Carbon atoms that do not belong to the hexagonal faces are marked in red color. Starting from an initial configuration (panel i) almost entirely tiled with hexagonal polygons, atoms rearrange to fill uniformly the surface (panels ii to v) reaching a local minimum after a few thousands steps (panel vi). After reaching this configuration, trial moves are rejected at an increasingly high rate and the Metropolis algorithm becomes inefficient. b) A dualization sequence (described in Supplemental Material \cite{supp}), leading to a thousandfold increase in the number of carbon atoms and close to a hundredfold increase in the pseudosphere radius.}
\end{figure*}
\indent Here we construct a solid-state black-hole analogue consisting of a graphene membrane characterized by a three-connected tessellation engineered to shape it in the form of a constant negative curvature surface, known as Beltrami's pseudosphere~\cite{Iorio:2011yz,Iorio:2013ifa,taioli2016}. In particular, we develop a novel computational method to build realistic and  energetically stable negative curvature carbon allotropes comprising millions of atoms. Furthermore, we elaborate a tight-binding (TB) approach to calculate the local density of state (LDOS) for these extended curved structures. Comparison between the numerically evaluated LDOS and the theoretically predicted one shows, within uncertainties, its thermal nature, establishing the presence of a black hole type horizon in the system.


Beltrami's pseudosphere represents the hyperbolic counterpart of the regular sphere: it is a surface of revolution characterized by a constant negative Gaussian curvature $\kappa=-1/R_\mathrm{p}^2$, with $R_\mathrm{p}$ the pseudosphere radius. \\
\indent Under suitable boundary conditions, Gauss Bonnet's theorem shows that the existence of Stone-Wales (SW) defects with an excess of six heptagonal defects with respect to the pentagonal units \cite{taioli2016,tatti2016synthesis} is required to tile the pseudosphere with carbon atoms. Thus, the presence of six heptagonal shapes is imposed at the beginning and preserved by all the steps of the construction. In addition, Hilbert's theorem states that no analytic complete surfaces of constant negative Gaussian curvature can be embedded in $\mathbb{R}^3$, implying that the graphene pseudosphere cannot be complete. 
\\
\indent
Early investigations to build a realistic  Beltrami's pseudosphere by finding a (local) minimum energy tiling of carbon atoms taking into account these two theorems~\cite{taioli2016}, have been inconclusive in: i) delivering a general approach to the tessellation of hyperbolic surfaces; ii) scaling-up the graphene pseudosphere size; and iii) measuring the surface's electronic structure. And properties ii) and iii) are of paramount importance in ascertaining the capacity of this carbon--based structure to act as an analogue gravity model. 
\\
\indent
Our method proceeds as follows: we start the pseudosphere generation from a planar graphene sheet, in which we impose the presence of six heptagonal faces in the center (see Fig.~\ref{fig:1}a-i). The initial configuration of the pseudosphere (Fig.~\ref{fig:1}a-ii) is then obtained by simply projecting the graphene net on the Beltrami's surface along the $z$-axis (see Supplemental Material \cite{supp}). In this configuration, the carbon-to-carbon bond lengths in the bent region within the pseudosphere is longer than the typical bond distances in flat graphene ($a_\mathrm{CC}=1.42$ \AA), owing to the (negative) curvature. Next, a sequence of bond-switching trial moves and structural optimization steps with a modified Keating potential to favour the formation of hexagonal cells is then applied (see Supplemental Material \cite{supp}, Fig.~5a), and  accepted or rejected according to a suitable energy minimization criterion (Fig.~\ref{fig:1}a, panels iii through v). After ${\cal O}(10^4)$ moves the algorithm efficiency drastically drops, which limits the radius size of the minimized structures (Fig.~\ref{fig:1}a vi) to few nm and the number of carbon atoms to ${\cal O}(10^3)$. \\
\indent Scaling-up of the numbers of atoms to achieve satisfactory experimental conditions (which will be discussed below) is next implemented through a custom dualization algorithm (Figs.~\ref{fig:1}b and 5b), by which the pseudosphere radius and number of atoms scale like $\sim\sqrt{3}$ and $\sim3$ respectively, while conserving both the bond distance as well as the number of defects (see Supplemental Material \cite{supp}). Each dualization step is then followed by a bond switching optimization run to counteract the former tendency of splitting apart the SW defects of the original structure (and, thus, artificially increasing its total energy). Repeated application of this procedure allows one to reach a thousandfold increase in the number of carbon atoms (our maximum value being $N=2,615,976$) and a pseudosphere radius $R_\mathrm{p}=73.96$ nm. 
We hasten to emphasize that these atomic configurations are found to be  stable also by molecular dynamics simulations at several thousands K. More specifically, despite the formation of ripples and local deformations in proximity of the defected sites, graphene membranes of minimal energy result dynamically stable also when relaxing the condition that carbon atoms are strictly located on the analytical Beltrami's surface.


\begin{figure}[!t]
	\includegraphics[scale=0.5]{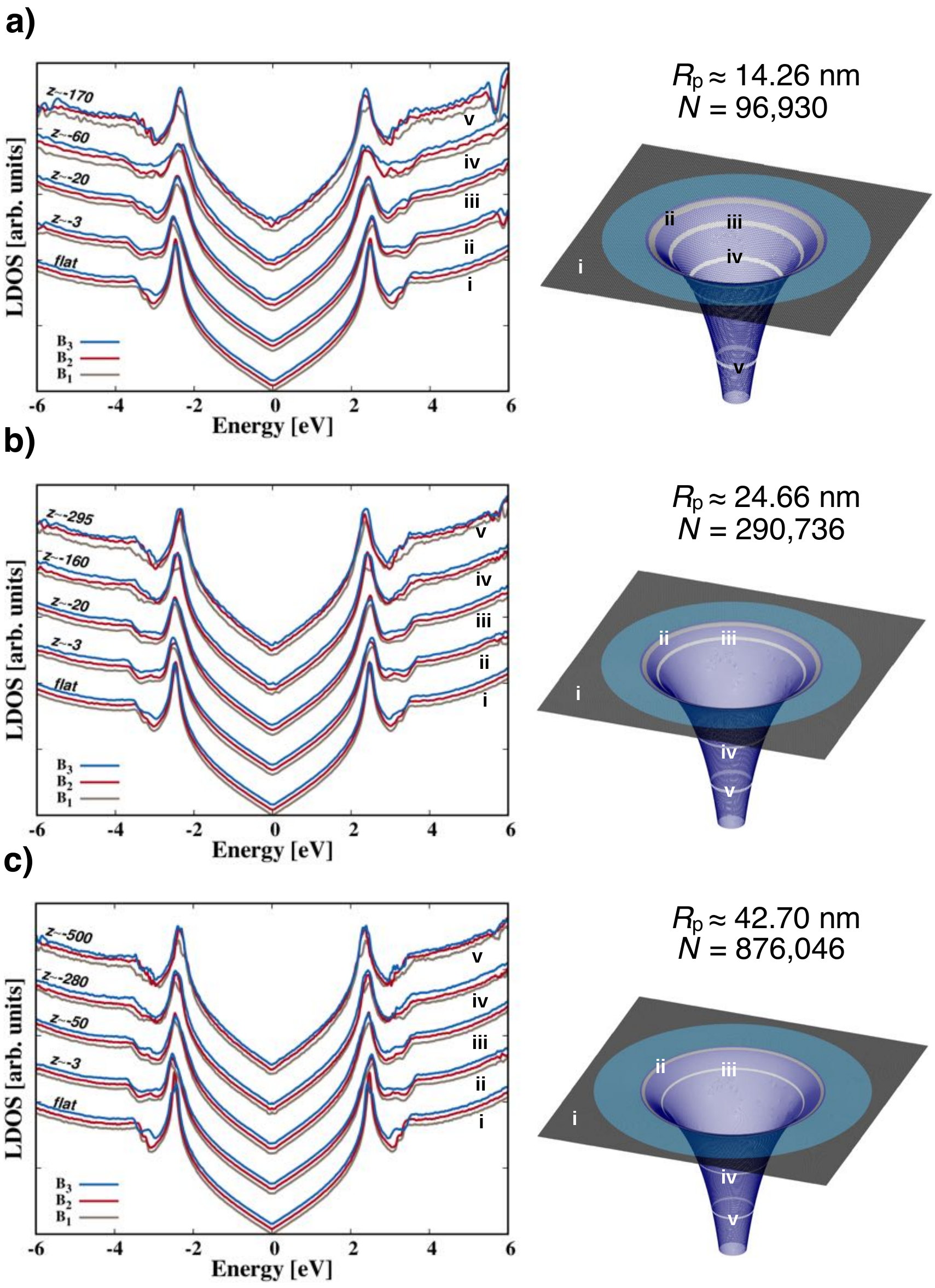}
	\caption{\label{fig:2}Evaluation of the LDOS through a multi-orbital TB approach implementing the Kernel Polynomial Method. Panels a) through c): LDOS projected onto five regions at different depth along the revolution axis $z$ (i through v) for various pseudospheres characterized by three different values of the number of atoms $N$ and radius $R_\mathrm{p}$. For each case of $N$ and $R_\mathrm{p}$ we report the LDOS for three pseudosphere realizations B$_{1\mathrm{-}3}$ differing by the configurations of SW defects, of which a representative is shown on the right of each panel. The Fermi energy is set equal to zero in all cases.}	
\end{figure}

The signature of the Hawking-Unruh effect in the carbon pseudosphere can be found by characterizing the electronic properties in terms of the LDOS near the Dirac points~\cite{neto2009}, where electrons behave as relativistic massless pseudo-particles. Given the ${\cal O}(10^6)$ carbon atoms of the realized structures, the LDOS will be evaluated through a multi-orbital TB approach implementing the Kernel Polynomial Method (KPM) to avoid the diagonalization of the Hamiltonian \cite{torres2014}. Due to curvature, in fact, the $p_z$ orbitals contributing to the $\pi$ band are not anymore orthogonal to the in-plane direction; similarly, the $sp^2$-hybridized orbitals do not lay in the graphene plane. Thus, an approach, in which all four valence orbitals ($2s,2p_x,2p_y,2p_z$) are included in the simulations as opposed to the $p_z$ orbital alone, has been necessary (see Supplemental Material \cite{supp} and Fig.~6 therein, where details concerning the parametrization of the Hamiltonian are reported and well-established results on graphene and carbon nanotube structures reproduced). 

\begin{figure}[!t]
	\includegraphics[scale=0.5]{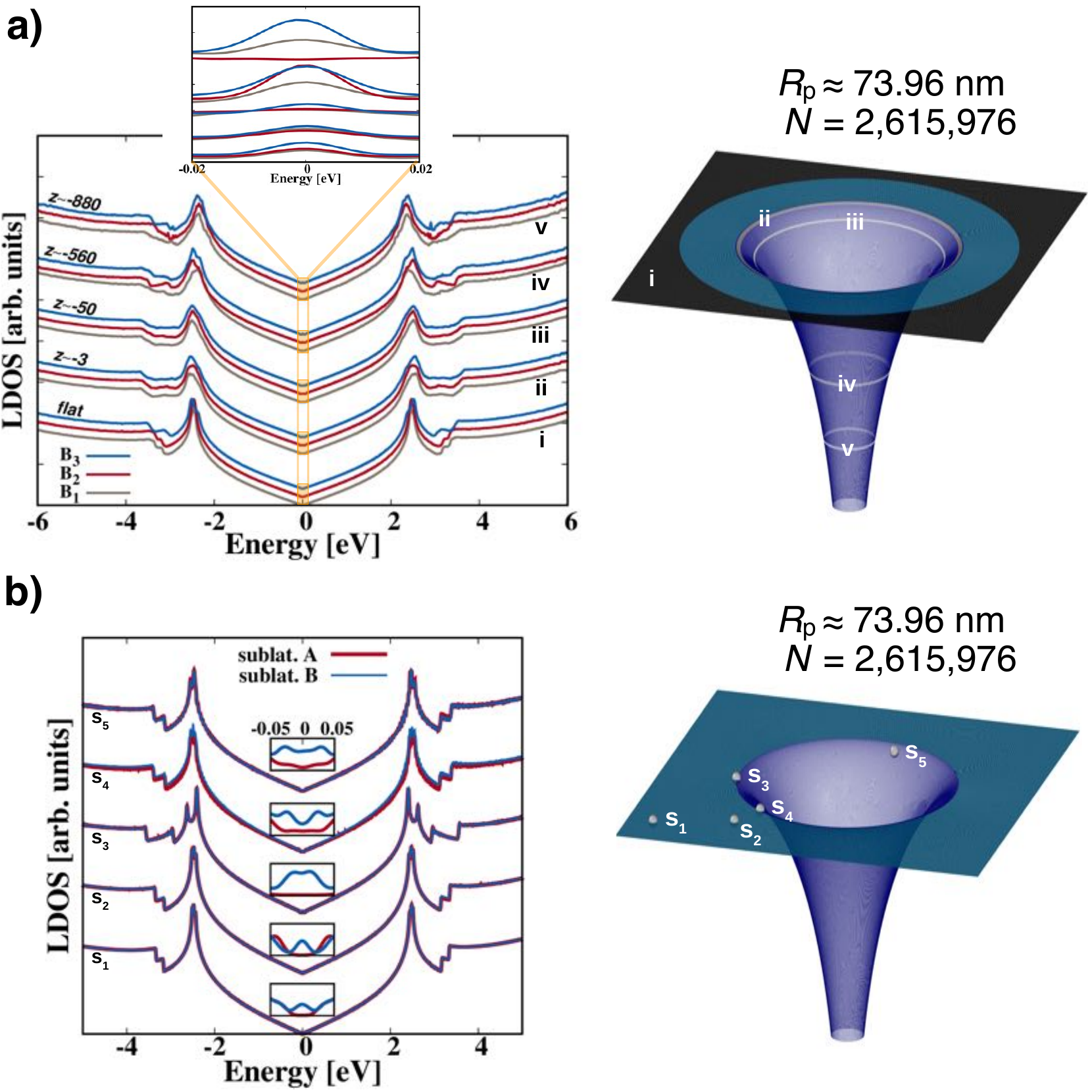}
	\caption{\label{fig:3}LDOS symmetry breaking due to curvature effetcs. a) LDOS projected onto five regions at different depth along the revolution axis $z$ (i through v) for for the three biggest pseudospheres ($N=2,615,976$ and $R_\mathrm{p}=73.96$ nm, right). We also zoom on the Fermi energy to expose the LDOS bulge in this region. b) Pseudosphere's LDOS projected over the shown sites $s_i$. Projection over the two inequivalent sublattices A and B of graphene is also shown; the insets zoom near the Fermi energy to show the LDOS asymmetry.}
\end{figure} 

\indent The LDOS projected onto longitudinal circles in regions located at a different $z$-depth along pseudospheres obtained at various stages of the dualization procedure (and thus characterized by varying number of atoms $N$ and radius $R_\mathrm{p}$), is plotted in Fig.~\ref{fig:2}. In each case we evaluate this quantity for three structures differing by the number and location of the SW defects. In the energy range $E\in[-6,6]$ eV, the LDOS shows a graphene-like shape for all the pseudospheres independently of the radius and defect distribution. With respect to the pristine graphene (region i), region ii shows Van-Hove singularities associated to the $\pi$ band peaks which are broadened and shifted; this is due to the slightly elongated carbon bonds characterizing this pseudosphere region, which represents the would-be Hilbert horizon (where the pseudosphere ends as a consequence of the Hilbert theorem). Moving further inside, the LDOS stays the same at a qualitatively level independently of the pseudosphere funnel depth at which is evaluated. 
\\
\indent
A similar overall behaviour (Fig.~\ref{fig:3}a) persists in the biggest structures studied. However, by zooming in the vicinity of the Fermi energy we find a bulge, which can be seen in the blown up region in Fig.~\ref{fig:3}a) and which could be also spotted in the central region of~Fig.~\ref{fig:2}. We attribute this behavior of the LDOS to a genuine curvature effect. Projecting the LDOS over single atomic sites both inside and outside the pseudosphere (sites $s_{1\mathrm{-}5}$ in Fig.~\ref{fig:3}b) and disentangling the nearest--neighbour contributions (that would correspond to the A and B sublattices in pristine graphene), we find that the LDOS spectrum around the Fermi energy $E_\mathrm{F}$ is significantly asymmetric for the two nonequivalent sublattices, while for energy $E\gg E_\mathrm{F}$ it is practically indistinguishable.


\begin{figure}[!t]
	\includegraphics[scale=0.49]{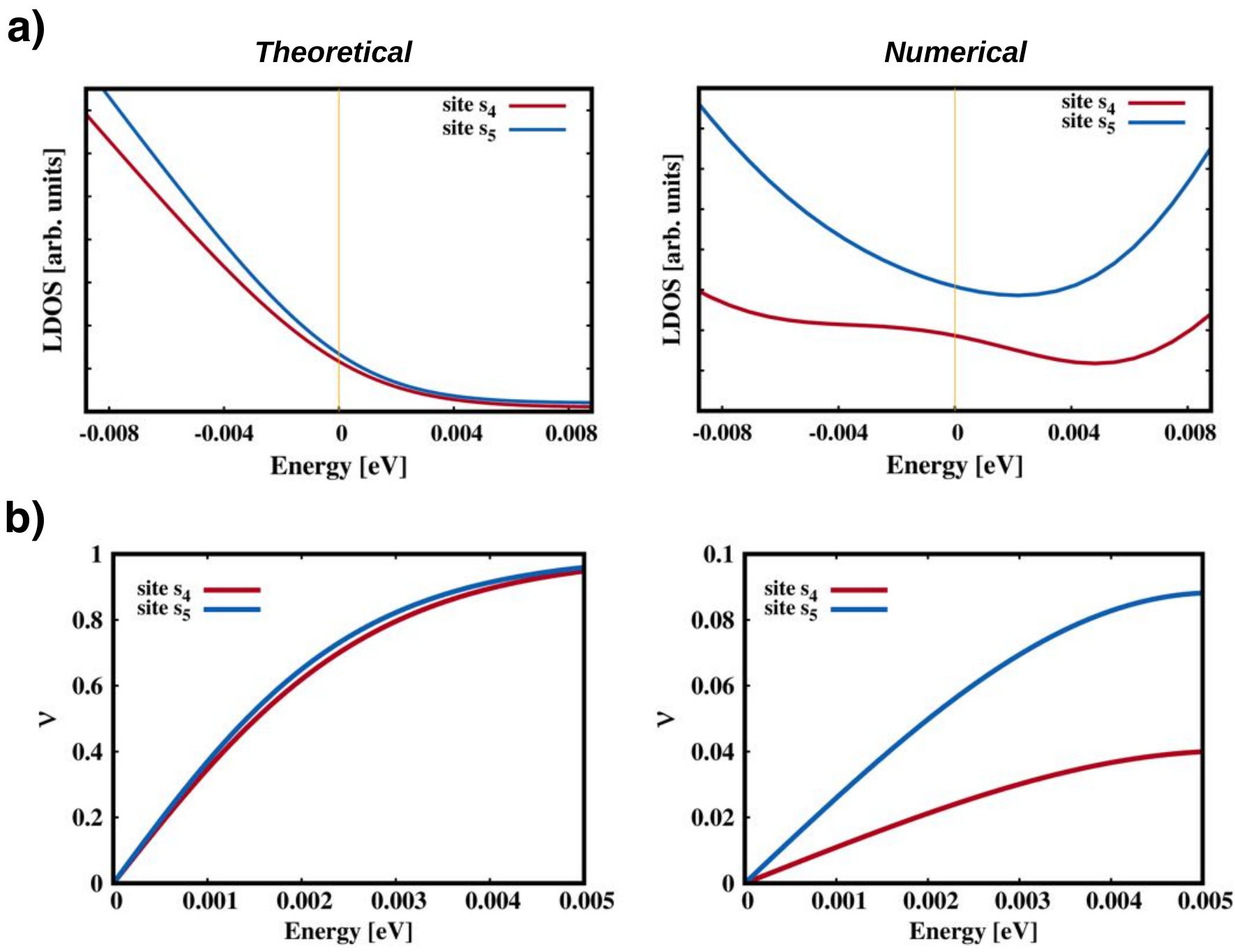}
	\caption{\label{fig:4}Detecting the presence of an event horizon through the low-energy LDOS. LDOS  lineshape (a) and contrast (b) projected over the pseudosphere sites $s_{4,5}$ shown in Fig.~\ref{fig:3}b), according to analytical predictions~\cite{Iorio:2011yz,Iorio:2013ifa} (left) and numerical results (right).}
\end{figure}

The largest simulated pseudosphere has a ratio $a_\mathrm{CC}/R_\mathrm{p}\sim2\times10^{-3}$. This parameter determines how well the Hilbert horizon and the Rindler-type event horizon, emerging when treating the pseudosphere as a 2+1 dimensional space-time in which the valence electrons move, coincide~\cite{Iorio:2013ifa} (0 representing coincidence). In the ideal case~\cite{Iorio:2011yz,Iorio:2013ifa}, in the proximity of the horizon, Hawking's radiation happens due to massless electrons, which  before tunnelling (i.e. on the pseudosphere surface) are described by the action
\begin{align}
	S=iv_\mathrm{F}\!\int\!\mathrm{d}^3x\,\sqrt{g}\bar\psi\gamma^\mu{\cal D}_\mu\psi,
	\label{action}
\end{align} 
where $v_\mathrm{F}\sim c/300$ is the Fermi velocity, $\gamma^\mu$ are the Dirac matrices, $\psi$ and $\bar\psi$ are the field operators creating particles and holes respectively, and ${\cal D}_\mu$ is the SO(2,1) covariant derivative. Finally, $g$ is the determinant of the pseudosphere metric $\mathrm{d}s^2_\mathrm{B}=\phi^2(u)\mathrm{d}s^2_\mathrm{R}$, where $\phi(u)=\ell/r\ \mathrm{e}^{u/r}$, $\ell$ is a constant that in the physical case is to be identified with $a_{\mathrm{CC}}$) and $\mathrm{d}s^2_\mathrm{R}$ is the Rindler-type metric $\mathrm{d}s^2_\mathrm{R}=\phi^{-2}(u)(\mathrm{d}t^2-\mathrm{d}u^2)-r^2\mathrm{d}v^2$ (with $u$ and $v$  the curvilinear coordinates spanning the pseudosphere). After tunneling, the electrons move in a flat metric (the graphene plane) where the action is given by Eq.~(~\ref{action}), with the replacements ${\cal D}\to\partial$, and $g\to1$. The presence of an event horizon, can be revealed by evaluating the power spectrum $\rho$ of the 2-point function $S_\mathrm{B}=\mbox{}\langle0_\mathrm{M}\vert\psi_\mathrm{B}\bar\psi_\mathrm{B}\vert0_\mathrm{M}\rangle$, being $|0\rangle_\mathrm{M}$ the flat vacuum; in this case, it would assume a thermal form, as neglecting boundary terms~\cite{Iorio:2011yz,Iorio:2013ifa}:
\begin{align}
    \rho&=\frac{4}{\pi}\frac{1}{(\hbar v_\mathrm{F})^2}\frac{R_\mathrm{p}^2}{a_{\mathrm{CC}}^2}\mathrm{e}^{-2u/R_\mathrm{p}}\frac{E}{\mathrm{e}^{\frac{E}{k_\mathrm{B}\Theta}}-1};&
    \Theta=\frac{\hbar v_\mathrm{F}}{k_\mathrm{B}}\frac{a_\mathrm{CC}}{2\pi r^2} \mathrm{e}^{u/r},
	\label{rhoil}
\end{align}
where $k_\mathrm{B}$ is the Boltzmann constant and $\Theta$ the temperature. At the horizon, where $r=R_\mathrm{p}$ and $u=R_\mathrm{p}\ln R_\mathrm{p}/a_\mathrm{CC}$, the Hawking temperature reaches its maximum $\Theta=\hbar v_\mathrm{F}/(2\pi\kappa R_\mathrm{p})\sim 16$ K for our largest pseudosphere. Notice that this is a low energy effect: only electrons with an intrinsic energy $E=\hbar v_\mathrm{F}/R_\mathrm{p}\sim9$ meV have a wavelength long enough to experience the effects of the curvatures and thus a LDOS described by~(\ref{rhoil}). Furthermore, the large radius requirement implicit in the intrinsic energy scale emerging from the graphene pseudosphere analog model rigorously justifies the low-energy description in terms of massless electron and holes. On the other hand, for detecting experimentally the Hawking temperature associated to the existence of the Rindler horizon, $R_\mathrm{p}$ should be not extremely large, as Eq.~(\ref{rhoil}) implies $\Theta \propto 1/R_p$. Thus, the optimal radius value turns out to be a trade-off between these two opposite requirements. We notice that already for $a_\mathrm{CC}/R_\mathrm{p}\sim10^{-2}$ (see Fig. \ref{fig:2}) the approximation of the Rindler event horizon with the Hilbert horizon of the Beltrami's spacetime is rather accurate and the LDOS asymmetry is emerging. Nevertheless a pseudosphere with a radius in the range of $\mu$m is necessary to achieve a good resolution in the linear part of spectrum.\\
\indent Given the exponential nature of the power spectrum~(\ref{rhoil}), the effect of the presence of the horizon should manifest in a marked asymmetry of the LDOS around the Fermi energy, as measured by the contrast
\begin{align}
	\nu(E)=\left\vert\frac{\mathrm{LDOS}(-E)-\mathrm{LDOS}(E)}{\mathrm{LDOS}(-E)+\mathrm{LDOS}(E)}\right\vert.
\end{align}
Results for the LDOS and its contrast projected on sites located near the Hilbert/Rindler horizon are shown in Fig.~\ref{fig:4} for the theoretical (left) and numerical (right) predictions. The behavior of the curves is remarkably similar, once we take into account that the continuum approximation formula~(\ref{rhoil}) is valid when neglecting both local elastic effects induced by the curvature as well as the presence of SW defects; and that in our realistic model, the interplay between curvature and defects (which the Gauss-Bonnet theorem makes inseparable, if not topologically equivalent) is inherently present and, thus, its effect on the LDOS of the graphene sublattices can be neither avoided nor disentangled. In particular, while being off in magnitude, the contrast is qualitatively the same.\\
\indent In summary, the Beltrami's pseudosphere tiled by carbon atoms arranged in a defected graphene net, which is found to be an energetically and dynamically stable allotrope of carbon, may represent a viable analogue model of a quantum field theory in curved space-time in general, and a black-hole horizon in particular. Massless electron-hole pair generation at the Hilbert horizon of the graphene pseudosphere as measured by the LDOS is analogous to Hawking radiation in conventional black holes; but while in the latter systems the radiation temperature is too small to be observed directly, in the carbon pseudosphere temperatures of the order of tens of K are in principle attainable.
The success of our numerical efforts in verifying the analytical predictions obtained within a continuum representation suggests that this analogue system, if experimentally realized in a lab (for example, through direct optical forging~\cite{Johansson:20173D}), can allow measuring a Hawking temperature several orders of magnitudes higher than the one detected in sonic analogues, by ascertaining the thermal character of the low energy LDOS through either low temperature scanning tunnelling microscopy or optical near-field spectroscopy. \\
\indent 
Finally, the structural model developed here could also represent a new paradigm to investigate other aspects of black hole physics such as in particular the modification of Hawking radiation signals in situation of close proximity of two black holes.

\indent N.M.P. is supported by the European Commission under the Graphene Flagship Core 2 grant No. 785219 (WP14, ``Composites'') and FET Proactive (``Neurofibres'') grant No. 732344 as well as by the Italian Ministry of Education, University and Research (MIUR) under the ``Departments of Excellence'' grant L.232/2016, the ARS01-01384-PROSCAN and the PRIN-20177TTP3S grants. The authors acknowledge Bruno Kessler Foundation (FBK) for providing unlimited access to the KORE computing facility.

%


\clearpage

\onecolumngrid
\section*{\LARGE SUPPLEMENTAL MATERIAL}
\section*{\large Exploring Event Horizons and Hawking Radiation through Deformed Graphene Membranes}
\subsection*{Tommaso Morresi,$^{1,2}$ 
Daniele Binosi,$^{1}$ 
Stefano Simonucci,$^{3}$
Riccardo Piergallini,$^{3}$ 
Stephan Roche,$^{4,5\ast}$
Nicola M. Pugno,$^{2,6,7}$ 
Simone Taioli$^{1,8\ast}$}
\noindent
$^{1}$European Centre for Theoretical Studies in Nuclear Physics and Related Areas (ECT*-FBK), Trento, Italy\\
$^{2}$Laboratory of Bio-Inspired \& Graphene Nanomechanics -- Department of Civil, Environmental and Mechanical Engineering, University of Trento, Italy\\
$^{3}$School of Science and Technology, University of Camerino, Camerino, Italy\\
$^{4}$ Catalan Institute of Nanoscience and Nanotechnology (ICN2),
CSIC and BIST, Campus UAB, Bellaterra, Barcelona, Spain\\
$^{5}$ICREA, Institucio Catalana de Recerca i Estudis Avancats, Barcelona, Spain  \\
$^{6}$Ket-Lab, Edoardo Amalfi Foundation, Rome, Italy\\
$^{7}$School of Engineering and Materials Science, Queen Mary University of London, UK \\
$^{8}$Trento Institute for Fundamental Physics and Applications (TIFPA-INFN), Trento, Italy\\~\\

\twocolumngrid

\subsection*{Tiling the pseudosphere}

\begin{figure*}[!t]
	\includegraphics[scale=0.68]{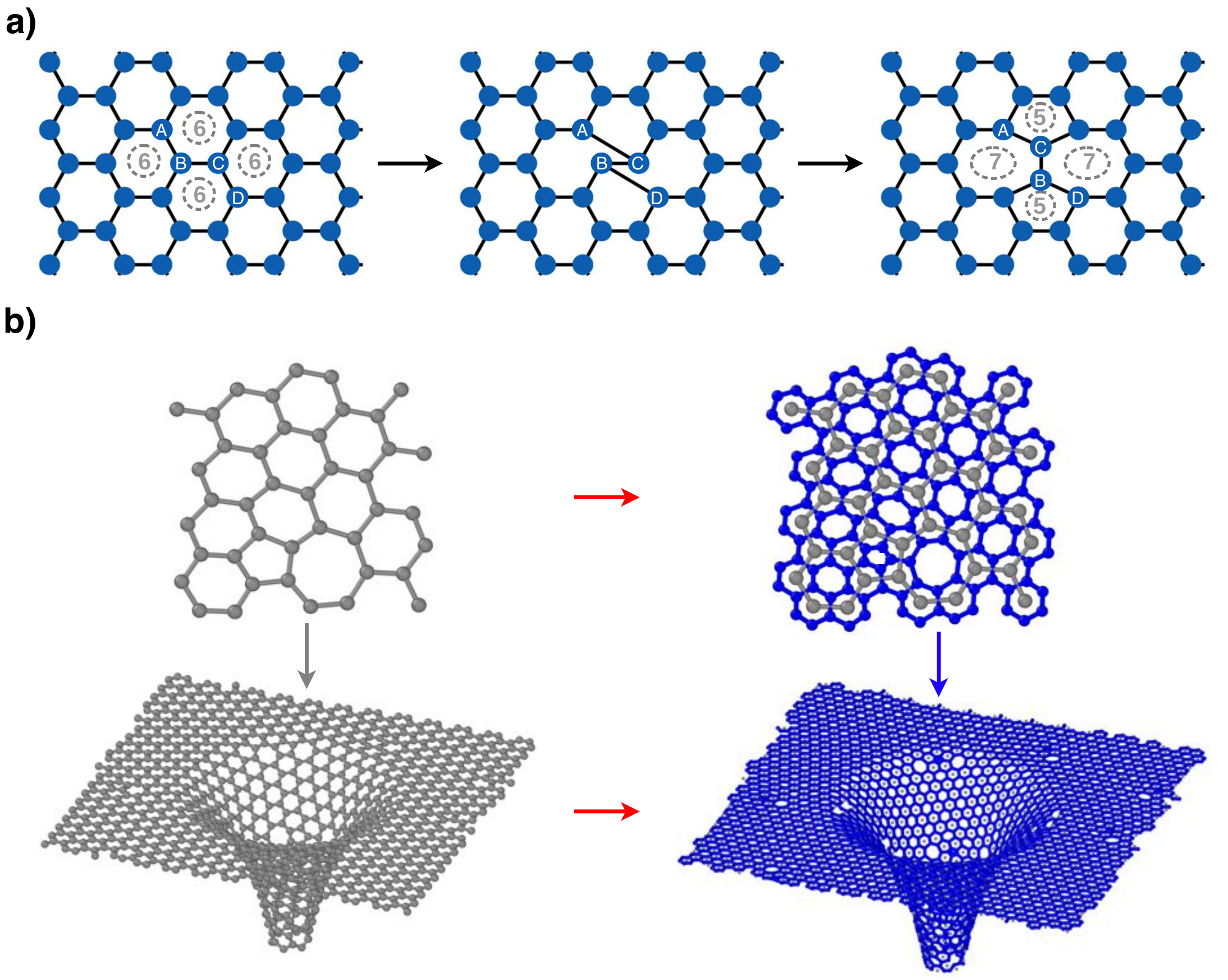}
	\caption{\label{fig:M1}Bond switch and dualization procedures. a) An example of a bond-switch trial move in a graphene lattice. Starting from an initial configuration showing 6-fold rings (dashed circles), the bonds between the carbon atoms A--B, and C--D are cut, and new bonds A--C and B--D are formed. Twisting then transforms the four 6-fold rings into two 5-fold rings and two 7-fold rings (adapted from Ref. \cite{alfthan2006}). b) Dualization sequence based on the three-connectivity of the graph. The parent geometry is shown on the left, while the resulting (daughter) geometry is on the right.}
\end{figure*}

The tiling of the Beltrami's pseudosphere by carbon atoms, which represents an interesting geometrical problem in its own right, has been achieved through the following steps:

\begin{enumerate}
\item Set the length of the pseudosphere by fixing the maximum value of the coordinate along the axis of revolution ($z$).
\item Determine the number of carbon atoms $N$ that are needed if one were to tile the surface of the Beltrami's pseudosphere with the same density of planar graphene  ($0.379$ atoms/\AA$^2$). Periodic boundary conditions are applied by using a rectangular supercell repeated along the $x$ and $y$ directions to saturate the outer carbon atom bonds belonging to $r=R_\mathrm{p}$ and $z=0$ (the Hilbert horizon).
\item Construct a planar graph $(N,F,E)$ consisting of $N$ vertices, $F$ faces and $E$ edges. The $N$ vertices represent compressed carbon atoms with shortened carbon-to-carbon bond lengths, $a_\mathrm{CC}<$1.42 \AA; each vertex is linked to three nearest neighbours by edges (representing bonds) and is shared by three faces.
\item Map the initial graph onto the Beltrami's pseudosphere surface via a one-to-one transformation by which the revolution axis coordinate $z$ of the vertices is unambiguously determined $\forall \sqrt{x^2+y^2}<R_\mathrm{p}$ by fixing 
\begin{eqnarray}
	\qquad z&=&z(x,y)\nonumber \\&=&R_\mathrm{p} \left[ \sqrt{1-\frac{x^2+y^2}{R_\mathrm{p}^2}} - \mathrm{atanh}  \sqrt{1-\frac{x^2+y^2}{R_\mathrm{p}^2}}\ \right];\nonumber \\
\end{eqnarray}
\item Find the atomic arrangements with $N\sim {\cal O}(10^3)$ that minimize a surface potential energy of the Keating type~\cite{kumar2012}
\begin{eqnarray}    \qquad E=\frac{3}{16}\frac{\alpha}{a_\mathrm{CC}^2} \sum_{i,j} \Big( r_{ij}^2-a_\mathrm{CC}^2 \Big)^2&+&\nonumber \\ \frac{3}{8}\frac{\beta}{a_\mathrm{CC}^2}\sum_{i,j,k} \Big( \mathbf{r}_{ij} \cdot \mathbf{r}_{ik}+\frac{a_\mathrm{CC}^2}{2}\Big)^2&+&c_h \sum_{F_i}(|F_i|-6)^2, \nonumber \\
\end{eqnarray}
where $\alpha=25.88$ eV \AA~$^{-2}$ is the bond stretching force constant, $a_\mathrm{CC}=1.42$ \AA, $r_{ij}$ is the distance between atoms $i$ and $j$, and $\beta \sim \frac{\alpha}{5}$ is the bond--bending force constant. Finally, the last term favours the formation of hexagonal faces: $F_i$ labels the polygons of the net, $|F_i|$ is the number of vertices of the polygons, and ,finally, one has $c_h=0.35$ empirically. To reach the energy minimum we repeated the following steps, typically ${\cal O}(10^4)$ times: 
\begin{itemize}
	\item Perform random switchings/twists of atomic bonds, based on the Wooten, Winer and Weaire (WWW) method \cite{wooten1985} (Fig.~\ref{fig:M1}a); 
	\item Let the geometry relax through molecular dynamics simulations based on the Fast Inertial Relaxation Engine (FIRE) approach\cite{bitzek2006};
	\item Accept the move only if it lowers the total energy of the system according to the Metropolis algorithm \cite{gould1988}.
\end{itemize}

\item Execute on the minimized surfaces a dualization sequence, to increase the number $N$ of atoms and correspondingly the radius of the pseudosphere (Fig.~\ref{fig:M1}b). By using the three-connectivity of the graph one creates a hexagon around each vertex of the initial optimized structure; rescale the bond lengths with a $\sqrt{3}$ factor and repeat from 5.
\end{enumerate}

\subsection*{Tight--binding parameter estimate}

\begin{figure*}[!t]
	\includegraphics[scale=0.56]{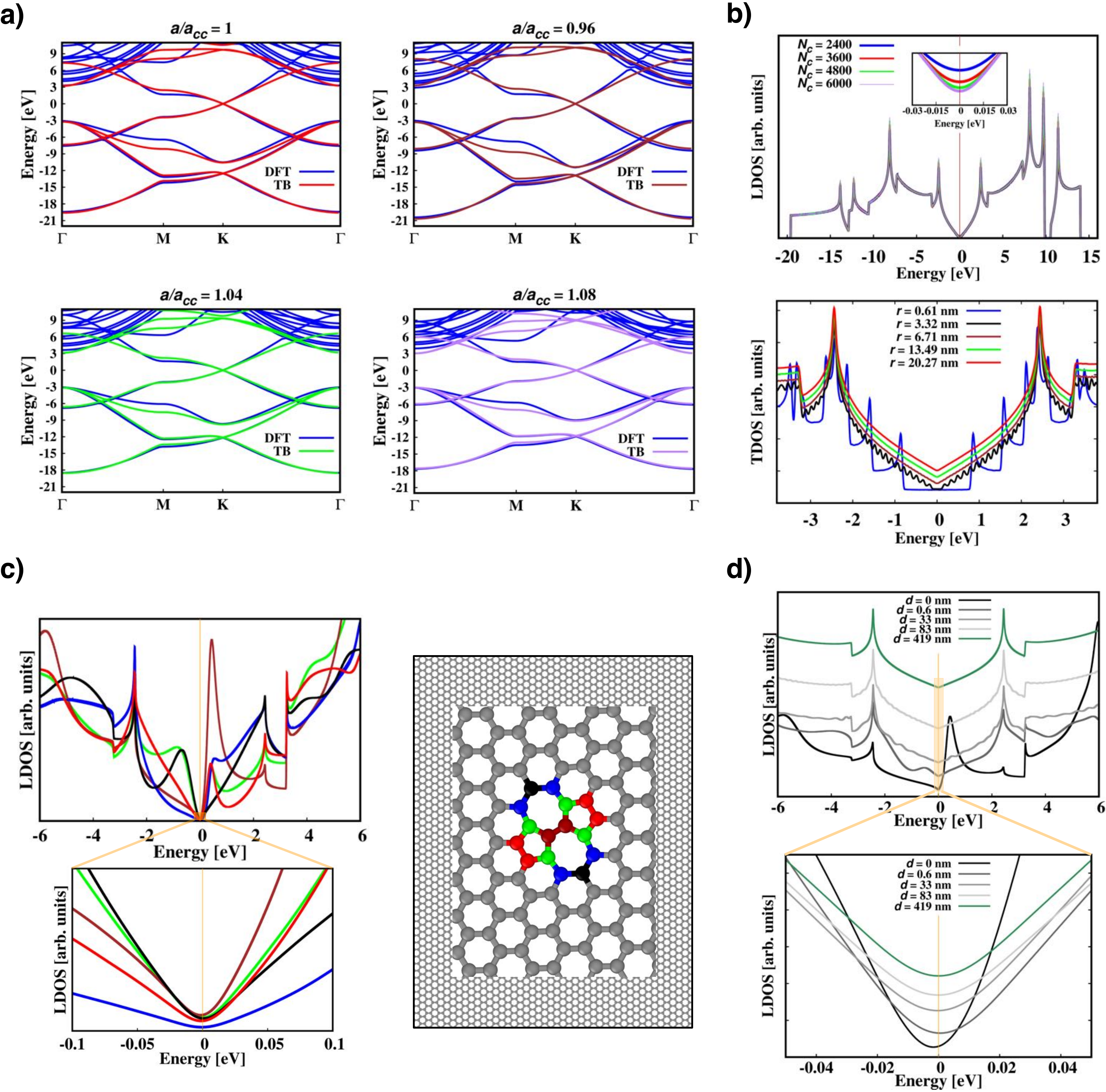}
	\caption{\label{fig:M2}Determination of the parameters for the multi-orbital TB algorithm and corresponding reference LDOS calculations. a) TB fit of graphene bands obtained from DFT electronic structure calculations. From left to right and from top to bottom we report the bands for a biaxial compression of the graphene cell equal to 4 \%, unstrained graphene, and  a biaxial tensile strain of the  cell equal to 4\% and 8\%, respectively. b) (top) LDOS for different values of the cut-off parameter $N_c$ obtained by projection on a site of a pristine  graphene rectangular cell with sides equal to $199.97$ nm $\times$ $199.81$ nm, containing $1,525,188$ carbon atoms; The inset represents a zoom near the Fermi energy. (bottom) total DOS of (n,n) nanotubes for n=9 ($r=0.61$ nm), $n$=49 ($r=3.32$ nm), n=99 ($r=6.71$ nm), n=199 ($r=13.49$ nm) and n=299 ($r=20.27$ nm). c) The LDOS angular dependence obtained by projecting over sites belonging to a SW defect (colored atoms in the right panel). d) The LDOS dependence on the radial distance $d$.}
\end{figure*}

Low energy electronic properties of geometries containing millions of atoms, have been evaluated using a TB approach, which is well known to describe correctly the dispersion of graphene around the six Dirac $K$-points in the first Brillouin zone~\cite{neto2009}. Due to the pseudosphere curvature, a multi-orbital TB approach has been developed, in which all four valence orbitals ($2s,2p_x,2p_y,2p_z$) are included in the simulation through the Hamiltonian:
\begin{equation}
	H=\sum_{\xi, i} \epsilon^i_{\xi} a^{\dagger}_{i,\xi} a_{i,\xi}+ \sum_{\xi,\gamma, \langle ij\rangle}t^{ij}_{\xi,\gamma} a^{\dagger}_{i,\xi} a_{j,\gamma},
	\label{tb}
\end{equation}
where $\xi,\gamma$ are orbital label indices while $i,j$ are site indices; $t^{ij}_{\xi,\gamma}$ indicates the hopping parameters; $a^{\dagger}$ and $a$ are the creation and annihilation operators; and the symbol $\langle ij\rangle$ means that the nearest neighbours approximation is adopted. The parameters $t^{ij}_{\xi,\gamma}$ describing the hopping between orbitals in different sites were computed within the Slater-Koster formulation \cite{slater1954}, which provides a scheme to relate the orbital symmetry, distances and directions of neighbour atoms. Owing to the non-planarity of our geometry we cannot make use of the multi-orbital parametrization typically used for graphene~\cite{yuan2015,stauber2016} where the onsite energy of the $p_z$-symmetry orbitals are treated differently from the $x,y$ orbital cartesian components along the in-plane directions (that is $\epsilon_{p_x}=\epsilon_{p_y} \ne \epsilon_{p_z}$). Therefore, we derive the TB parameters by fitting ab-initio Density Functional Theory (DFT) simulations of the graphene bands by further imposing that the onsite energies for the $p$ orbitals are the same ($\epsilon_{p_x}=\epsilon_{p_y} = \epsilon_{p_z}$). DFT simulations of equilibrium and strained configurations of graphene were carried out by using the Quantum Espresso code suite~\cite{qe}; in particular we use a norm--conserving PBE pseudopotential (C.pbe-mt gipaw.UPF) and an energy cut-off for the wavefuntion expansion on plane-waves set equal to 100 Ry. The $k$-point mesh is a $40 \times 40 \times 1$ grid for the calculation of both the ground state density and the band structures. Convergence of the integrals over the Brillouin zone was improved by smearing the occupancy with a 0.136 eV width Gaussian function. The TB parameters that we obtained using Eq. (\ref{tb}) for unstrained ($a = a_\mathrm{CC}$) and strained ($a \ne a_\mathrm{CC}$) graphene are: $\epsilon_s=-2.8$ eV and $\epsilon_p=0$ eV as onsite energies; $V_{ss \sigma}(a)=-5.6 \cdot a / a_\mathrm{CC} \cdot e^{-\frac{a-a_\mathrm{CC}}{0.55}}$ eV, $V_{sp \sigma}(a)=5.2 \cdot a/a_\mathrm{CC} \cdot e^{-\frac{a-a_\mathrm{CC}}{0.75}}$ eV, $V_{pp \sigma}(a)=4.6 \cdot a/a_\mathrm{CC} \cdot e^{-\frac{a-a_\mathrm{CC}}{0.55}}$ eV and $V_{pp \pi}(a)=-2.44 \cdot a/a_\mathrm{CC} \cdot e^{-\frac{a-a_\mathrm{CC}}{0.41}}$ eV as hopping parameters between different orbitals. In Fig.~\ref{fig:M2}a we report the bands of unstrained and strained graphene obtained by using the DFT and multi-orbital TB approaches.

\subsection*{Kernel Polynomial Method}

For the evaluation of the LDOS we resorted to the KPM, which is a numerical approach useful to access spectral quantities of extended systems for which a direct diagonalization of the full Hamiltonian matrix is computationally unfeasible. It consists in the expansion of the sought quantity in terms of a set of orthogonal polynomials, and then in improving the convergence of the expansion with a kernel to avoid spurious Gibbs oscillations~\cite{weisse2006}. In particular, we used the Chebyshev polynomials for the expansion, and the Jackson kernel to increase convergence, resolution, and accuracy~\cite{weisse2006}. Within this framework, a generic function can be expanded according to 
\begin{align}
	f(x) = \frac{1}{\pi \sqrt{1-x^2}} \left[ \mu_0 g_0+ \sum_{n=1}^{N_c-1} \mu_n g_n T_n(x) \right],
	\label{exp2} 
\end{align}
where $T_n(x)$ are Chebyshev polynomials of the first kind, $\mu_n=\int_{-1}^{1}\mathrm{d}x\,f(x)T_n(x)$ are the coefficients of the expansion and the $g_n$ are the Jackson kernel coefficients defined as
\begin{eqnarray}
    g_n&=&\frac1{N_c+1}\Big[(N_c-n+1) \cos\frac{\pi n}{N_c+1}\nonumber \\ &+&\sin\frac{\pi n}{N_c+1}\cot\frac{\pi}{N_c+1}\Big].
    \label{jakson}
\end{eqnarray}
Finally, $N_c$ represents the truncation number related to the maximum momentum. The best achievable resolution through this kernel is 
\begin{align}
    \Delta^J=\sqrt{1-\cos\frac{\pi}{N_c+1}}.
    \label{resolution}
\end{align}
We refer to~\cite{weisse2006} for the details about the calculation of $\mu_n$; here it suffices to emphasize that it is based on the stochastic evaluation of traces, which requires a certain number $R$ of random initial states. As expected, the bigger is $R$, the more accurate becomes the evaluation of the coefficients; we found that $R=100$ was enough for all calculations carried out.  
\subsection*{Tests of the LDOS calculations}

The convergence with respect to the $N_c$ parameter can be tested in the calculation of the LDOS for the benchmarks cases of planar graphene and armchair carbon nanotubes. The LDOS of graphene for four different values of $N_c$ ranging from 2400 to 6000, is shown in the top panel of Fig.~\ref{fig:M2}b). While at a wide energy scale the curves are indistinguishable, zooming near the Fermi energy (set to zero as usual) shows that higher truncation values for $N_c$ captures more faithfully the expected linear dispersion relation; on the other hand, there is a threshold to the number of terms in the summation after which spurious oscillations set in, thus spoiling convergence. This can be understood by noticing that the energy separation between levels in periodic graphene is infinitesimal and the DOS is a continuous function. Then, since the pseudosphere in our simulations is a large but finite system and the energy separation of the levels increases with respect to infinite periodic structures, a too big value of $N_c$ may result in a KPM energy resolution marginally above the finite energy separation between levels of our finite system, thus leading to poor convergence~\cite{garcia2015}. For non-planar  systems, we have computed the total DOS of (n,n) nanotubes for n=$9,\ 99,\ 199$ and $299$ (radius $r=0.61,\ 3.32,\ 6.71$ and $13.49$ nm). The total DOS is reported in the bottom panel of Fig.~\ref{fig:M2}b), where we observe that the DOS lineshape of these armchair nanotubes is reproduced surprisingly well already for the moderate value of $N_c=2000$ and that, as expected, the confinement effects become less important upon increasing the radius size. \\
\indent Since a method for estimating the value of $N_c$ that trades-off between accuracy and computational efficiency exists only for pristine structures that do not have any defect~\cite{garcia2015}, selecting the best $N_c$ is a trial and error process. For the pseudosphere case we found $N_c=8000$ to be the optimal value (with $N_c=16000$ used when resolving the Fermi energy region in Fig.~\ref{fig:4}).\\ 

\subsection*{LDOS of a single SW defect in planar graphene}

The SW defects are present within the realistic framework of the Beltrami's pseudosphere owing to the negative curvature, while their occurrence is neglected in the analytical continuum model. Thus, we finally investigate the effect of the presence of a single SW defect on the LDOS of a graphene net ($N=823,860$), particularly near the Fermi energy where our interest is focused. We study both the LDOS projected over different symmetry sites of the SW defect (with $N_c=5000$), thus obtaining information on the angular dependence (Fig.~\ref{fig:M2}c), as well as the LDOS projected over sites increasingly far from the SW defect, thus obtaining insights on the radial dependence (Fig.~\ref{fig:M2}d). On top of a marked angular dependence, we see that the shape of the LDOS is dramatically modified near the defect site, while far from it the planar graphene shape is recovered; the presence of the SW defect still affects the LDOS projected at distances of $\approx 80$ \AA~ with small oscillations in the spectrum. Furthermore, we notice most importantly that near the Fermi energy one observes a marked asymmetry of the LDOS spectrum, persisting again up to a distance of $\approx 80$ \AA~. This effect overlaps in this energy range and actually is indistinguishable from the asymmetry owing to the negative curvature.

\end{document}